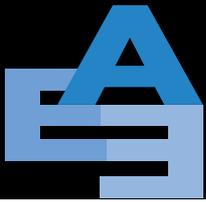
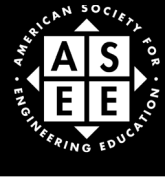

**Advances in Engineering Education**

# VIEW – A Virtual Interactive Web-based Learning Environment for Engineering

PRIYA T. GOESER
and
WAYNE M. JOHNSON
Engineering Studies
Armstrong Atlantic State University
Savannah, GA

FELIX G. HAMZA-LUP
Computer Science and Information Technology
Armstrong Atlantic State University
Savannah, GA
and

DIRK SCHAEFER
Mechanical Engineering
Georgia Institute of Technology
Savannah, GA

## ABSTRACT

The use of computer-aided and web-based educational technologies such as Virtual Learning Environments (VLE) has increased significantly in the recent past. One example of such a VLE is Virtual Interactive Engineering on the Web (VIEW). VIEW is a 3D virtual, interactive, student-centered, framework of web-based modules based on the Extensible 3D standard. These modules are dedicated to the improvement of student success and learning. In this paper, an overview of the recent developments in VIEW along with associated assessment results is presented. An experimental study was also performed to compare the learning experience and performance of students in a physical dissection activity vs. that in a virtual dissection activity using a VIEW module. The results of this study show that students can meet given learning objectives and that there is limited difference in their learning and performance irrespective of a physical or virtual setting.

Keywords: virtual learning environments, web-based labs, virtual mechanical dissection

## INTRODUCTION

Many engineering programs throughout the United States face the challenges of improving student learning, increasing student success and maintaining and/or improving retention rates [1]–[4]. In order to address these issues, educators have developed and implemented several pedagogical approaches to teaching and learning including project-based learning, student-centered learning, and computer-based learning, etc., that are well grounded in the scholarship of education and instructional techniques [5]. In addition, an increasing number of people are participating in online education and distance learning programs for various reasons and the demand for such programs is growing [6], which leads to the prevalent use of computer-aided and web-based educational technologies in the educational market. This includes the so-called Virtual Learning Environments (VLEs) [7], [8]. One example of such a web-based VLE is *Virtual Interactive Engineering on the Web (VIEW)*, which is a set of 3D web-based modules based on the Extensible 3D (X3D) standard. X3D can be used to develop applications that enhance student learning across a broad range of applications including computer-aided design (CAD), visual simulation, medical visualization, and geographic information system (GIS).

The modules in VIEW provide a student-centered, interactive, and engaging learning environment and are currently used in freshmen and sophomore engineering courses at Armstrong Atlantic State University (AASU). In general, VIEW may be employed as a supplement in traditional teaching and learning paradigms, as well as in remote laboratories, that is, physical laboratory experiments that are remotely controlled through the internet and can be taken by distance learning students any time and from anywhere with an internet connection [9]–[11].

In the following sections the authors present an overview of VIEW, background information regarding its underlying technology: the Extensible 3D standard, details on two fully developed modules of VIEW, associated assessment results, a study of physical vs. virtual learning environments, concluding remarks as well as an outline of future work.

## OVERVIEW OF VIEW

Virtual Interactive Engineering on the Web (VIEW) is a 3D virtual, interactive, student-centered, framework of web-based learning modules based on X3D and dedicated to the improvement of student learning, recruitment and retention in engineering programs. After its development, VIEW has been used at AASU since Fall 2008 with the following specific teaching and learning aims in mind:

1. Increase student exposure and understanding of how the components of mechanical mechanisms interact with each other to convey motion in various devices and systems.

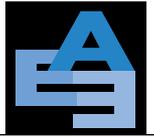

2. Improve student understanding of engineering material property characteristics and performance.

3. Improve student spatial visualization skills as it pertains to material property structures and 3D parts used in design.

Currently, there are two phases of VIEW as shown in Figure 1: Engineering Materials and Mechanical Dissection. The phase on Engineering Materials contains the module: The Virtual Tensile Testing Laboratory and the phase on Mechanical Dissection houses a module on the Mechanical Assembly of a Power Toothbrush. These phases have been implemented based on the above aims. Further details of VIEW can be obtained from the following URL: http://projects.felixlup.info/view/ and references [12], [13].

The primary objective in this paper is to address the following scholarly issues:

1. Evaluate the effectiveness of virtual learning environments such as VIEW on student learning and student success.

2. Investigate if virtual dissection modules as implemented in VIEW provide students with an equivalent learning experience as physical dissection modules.

3. Investigate if virtual learning environments such as VIEW may provide students in distance learning settings with an equivalent learning experience to those in traditional settings.

### BACKGROUND: EXTENSIBLE 3D

In engineering education, practical laboratory experience is an essential part of the learning experience wherein students are introduced to data analysis, problem solving, testing, and scientific interpretation. However, such laboratories are not always available because some universities have limited resources in terms of space or funds. One emerging trend to overcome these issues is to develop web-based 3D virtual laboratories to supplement physical laboratories and/or to reinforce important concepts studied in the courses [14]–[17]. In addition, such web-based virtual laboratories can be used in concert with remotely controlled physical laboratory experiments that are anticipated to close one of the major gaps in distance learning education. Currently, there are a number of virtual engineering labs/tools in use at several universities. To mention just a few, a virtual laboratory at John Hopkins University includes experiments showing the diffusion process, etc. [18]; a virtual torsion laboratory has been proposed by Bhargava et. al and is used by the faculty and students at Cornell University [19] and the Instructional Remote Laboratory continues to be used at Rutgers University and the University of Illinois at Urbana Champaign [20]. Some of these labs are implemented based on Hyper Text Markup Language (HTML), combined with applets or Flash™ technology and often lack a guided interaction process to encourage student engagement. In an effort to develop web-based 3D modules with student interaction, the Extensible 3D (X3D) standard was selected to serve as the technological backbone for VIEW.

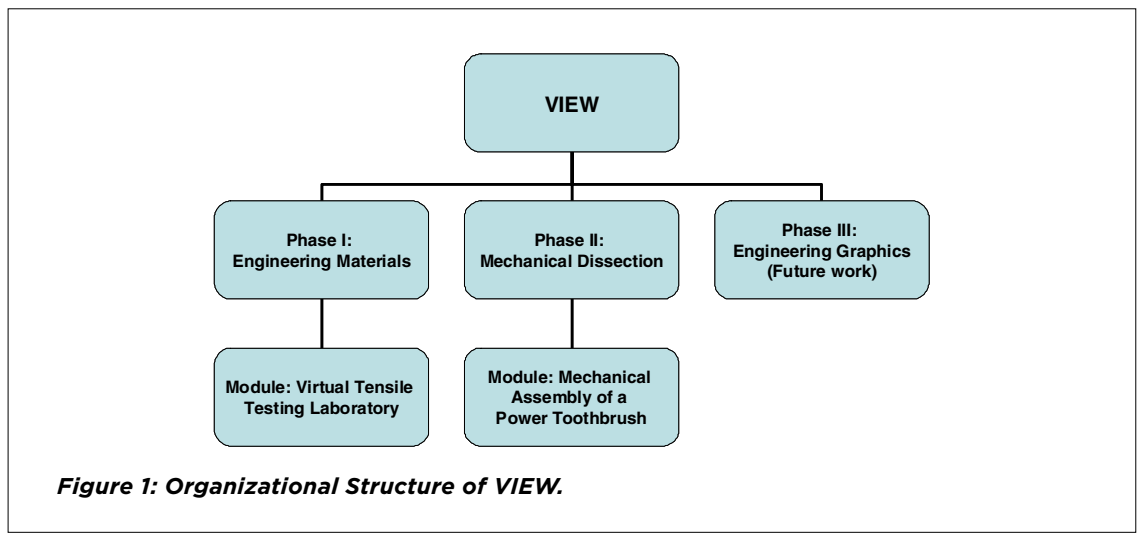

*Figure 1: Organizational Structure of VIEW.*

X3D is a scalable and open software standard for defining and communicating real-time, interactive 3D content for visual effects and behavioral modeling [21]. It can be used across hardware devices and in a broad range of applications including interactive simulations used for engineering education. X3D provides both the XML-encoding and the Scene Authoring Interface (SAI) to enable both Web and non-Web applications to incorporate real-time 3D data, presentations and controls into non-3D content. As a successor to the Virtual Reality Modeling Language (VRML), X3D is a more mature and refined standard [22]. Some additional features of X3D include [23]:

- Compatible with the next generation of graphics files, e.g. Scalable Vector Graphics.
- Open source (no licensing fees).
- Has been officially incorporated within the MPEG-4 multimedia standard.
- XML support makes it easy to expose 3D data to Web Services and distributed applications.
- 3D objects can be manipulated in C or C++, as well as in Java.

## THE VIRTUAL TENSILE TESTING LABORATORY

The Virtual Tensile Testing Laboratory (VTTL) was developed as a module in VIEW to be used as a supplement in the course: ENGR2000 - Introduction to Engineering Materials. This is a 3-credit hour lecture course taken as a required course by mechanical engineering sophomores and as an

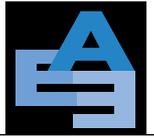

elective by civil and electrical engineering students. The course covers the fundamentals of ma-
terials processing, materials structure, material properties and testing, and materials performance
in various engineering applications. Students often view this course as a collection of abstract
concepts that are difficult to understand and relate to practical applications. This contributes to
the challenges to maintain a high level of interest, enthusiasm and information retention among
the students. The main objective of the VTTL is to introduce students to testing techniques such
as tensile testing, which can be used to evaluate certain mechanical properties of materials such
as the elastic modulus, yield strength, ultimate tensile strength, ductility, toughness and fracture
strength while in their sophomore year. This VTTL module is used as a supplement to the course in
order to enhance student learning of important concepts. Students continue to take the physical
tensile testing laboratory in their junior and senior years.

The VTTL consists of a realistic and detailed 3D CAD model of a tensile testing machine as shown
in Figure 2. The model was created using SolidWorks [24] and imported into the virtual scene us-
ing X3D. The graphical user interface (GUI) for the laboratory features HTML controls, a virtual 3D
scene, and a graph/display panel and is shown in Figure 3. The HTML controls include clickable
images allowing the user to choose a material sample and buttons for starting the experiment, ex-
amining the current sample in predefined views and downloading experimental data as is shown:
http://www.youtube.com/watch?v=sXB_6C3Ix2o. This data are then analyzed by the students to
determine relevant material properties. Further details on the implementation of the VTTL can be
found at the following reference [12].

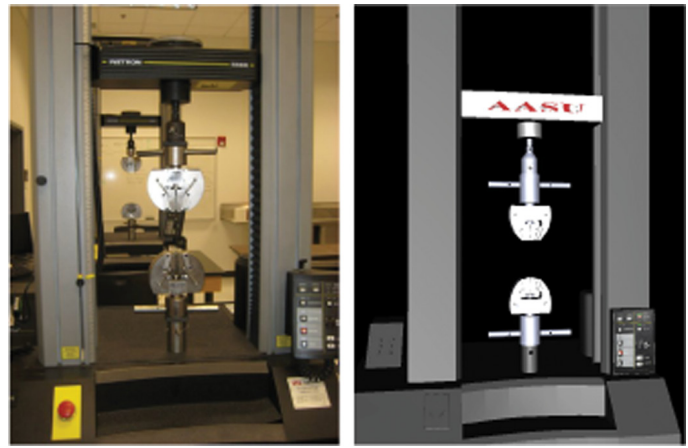

*Figure 2: InstronTM 5566 real TTM (left) and simulator (right).*

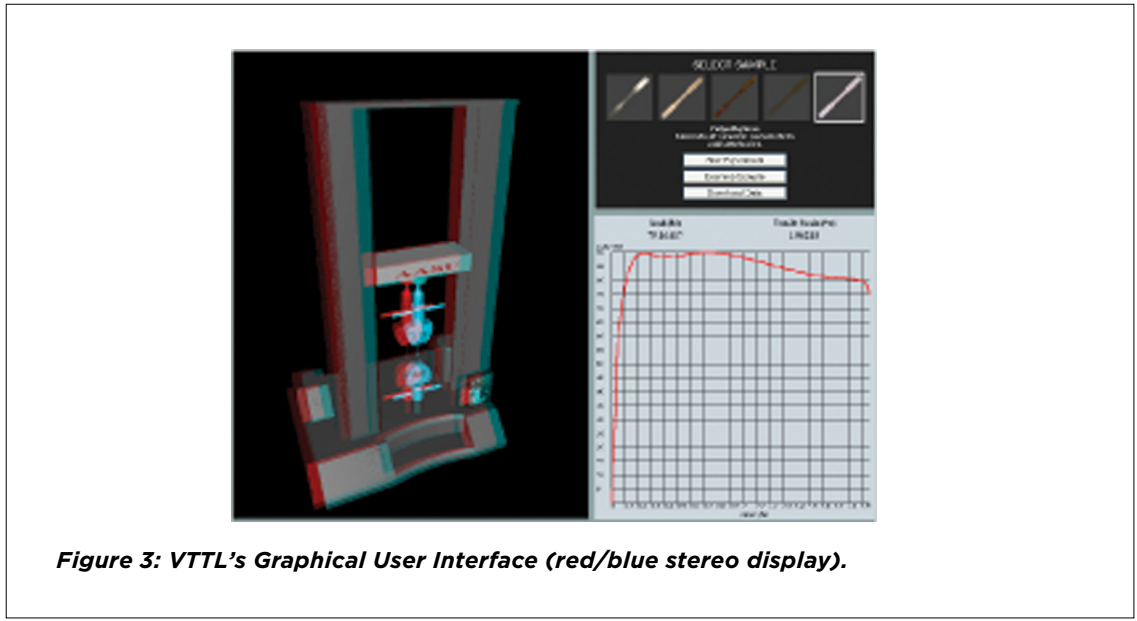

*Figure 3: VTTL's Graphical User Interface (red/blue stereo display).*

**Assessment**

A two-fold approach was used to assess the effectiveness of the VTTL on the students' learning experience. The first part consisted of evaluating students' perceptions of VTTL using a student survey, and the second compared student examination performance with and without the benefit of VTTL. The surveys were given to students to complete and return anonymously. Figures 4 and 5 show a summary of results of this survey for Fall 2008 and Fall 2009, respectively.

The bar graphs in Figures 4 and 5 correspond to the answers to the following questions:

**Question 1** (category Q 1): The virtual tensile testing laboratory helped me better understand the use of the stress-strain curve to determine mechanical properties of materials.

**Question 2** (category Q 2): The use of interactive, 3D models will help me better understand topics such as crystal models, miller indices, dislocations, etc. that require significant visualization skills.

**Question 3** (category Q 3): The use of virtual engineering labs/modules will help to supplement the need for 'hands on' projects and labs in the course.

**Question 4** (category Q 4): The use of virtual engineering labs/modules will be too time consuming and not very beneficial.

**Question 5** (category Q 5): The use of virtual engineering labs/modules in other engineering courses (ENGR1100, ENGR1170, ENGR2110, etc.) will help better understand topics that require significant visualization skills and/or 'hands on' projects and labs.

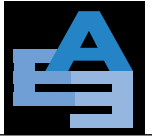

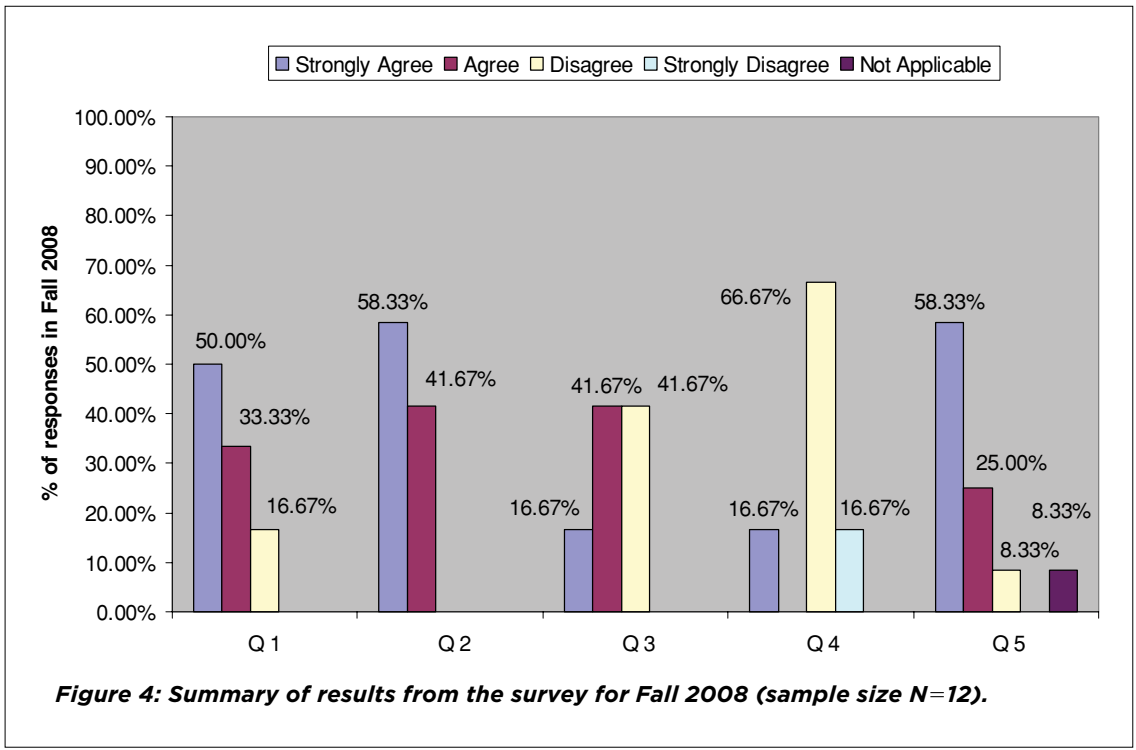

*Figure 4: Summary of results from the survey for Fall 2008 (sample size N=12).*

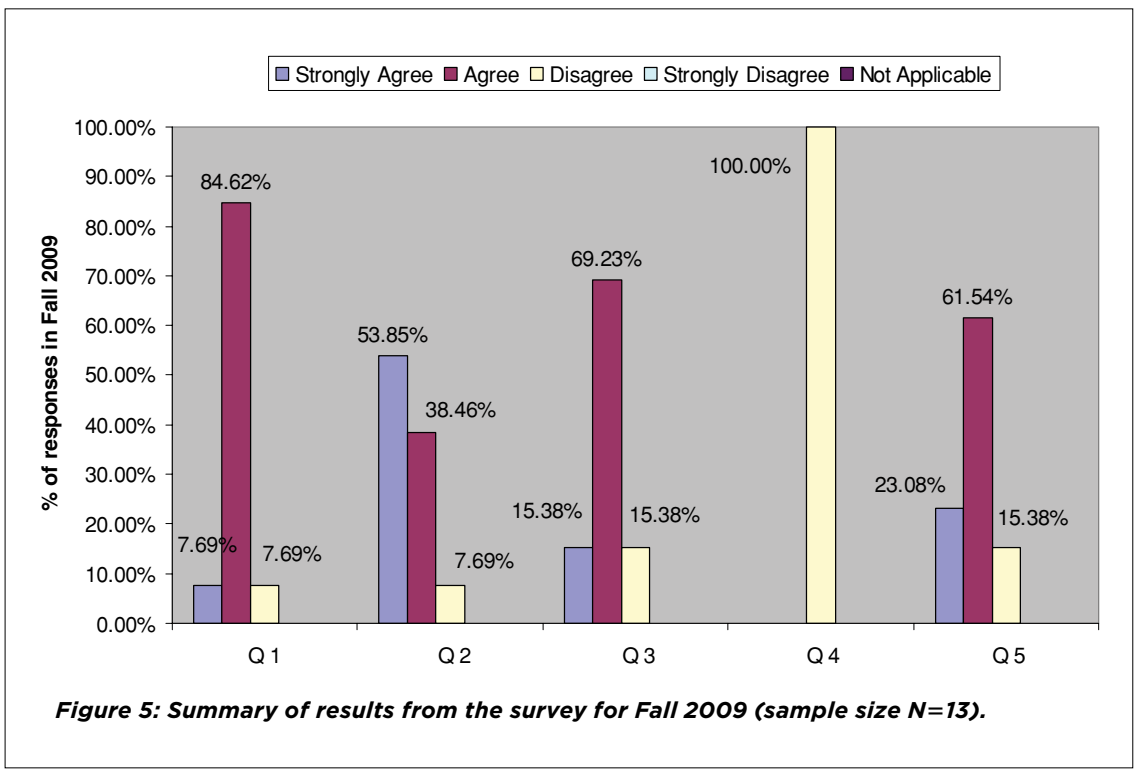

*Figure 5: Summary of results from the survey for Fall 2009 (sample size N=13).*

It is observed from the above data that 88% of the respondents strongly agreed or agreed that the virtual laboratory helped them to better understand the use of the stress-strain curve to determine mechanical properties of materials, an important concept studied in the course. A significant majority of the students also agreed that the use of interactive, 3D models will help them to better understand other topics relevant to this course. It is also observed that 41.67% (5 out of 12 students) in Fall 2008 and 15.38% (2 out of 13 students) in Fall 2009 say that the use of virtual engineering labs will NOT help to supplement the need for 'hands on' projects and labs in the course. This negative response can be attributed to the misconception that students perceive the VTTL module as their only exposure to such an engineering lab. The objective of the VTTL however, is to introduce students to the use of a tensile testing laboratory to determine material properties. Students continue to take the physical tensile testing laboratory in their junior and senior years.

Student's general comments are shown in Table 1. The majority of student comments were positive. Overall, the students agree and see the benefits of using interactive 3D labs/modules in this and other courses.

In order to assess student performance specific to the concepts and use of the tensile testing laboratory, a comparison of grades based on a problem (in the final exam) was done between the students who took the course in Fall 2007 (prior to the use of VTTL) and those who took it in Fall 2008 and Fall 2009 (after VTTL). A summary of these results is shown in Figure 6. It was observed that with the use of the VTTL in Fall 2008 and 2009, there was a 20.83% to 13.33% increase in the number of students (% of students enrolled) who received a grade of 75% or higher on this particular problem.

| Other comments | Number of similar responses |
|---|---|
| "I really liked the virtual lab – it provided the experience of computing with experimental data w/o the hassle of doing the physical experiment. I think ENGR1100 students would really enjoy seeing some realistic applications of engineering by using the virtual lab." | |
| "I really liked the simulated tensile test, however I would have liked to see better graphics illustrated in the type of fracture. Although I don't think that simulated labs could replace 'hands on' labs, I think it was a great way to review and apply all the material that we have learned so far." | 3 |
| "3D labs allow for better understanding for certain students who may be having trouble visualizing things from charts and figures. I think that more utilization of the 3D materials will be extremely beneficial." | 4 |
| "It is nice to see it in pictures as opposed to formulas" | |

*Table 1: Summary of other comments from students in Fall 2008 and Fall 2009.*

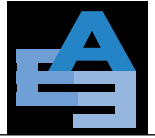

Based on the results of the surveys, student comments and student performance, the VTTL was well received by the students. In addition to benefits such as better understanding of concepts, students also learned the importance of technical writing skills that were required in writing the laboratory report.

## THE MECHANICAL ASSEMBLY OF A POWER TOOTHBRUSH

Mechanical dissection is an engineering activity that can satisfy a student's curiosity of how and why the components of given devices can convey specific motions to achieve a desired result [25]. Hence, several university engineering programs have developed mechanical dissection laboratories. However, such laboratories are not always feasible due to the lack of space, personnel, time and high costs. This issue is now being addressed through the use of multi-media technology to replace/supplement physical laboratories [26], [27]. Virtual dissection/assembly implemented in VIEW would only require the use of existing computer laboratories with an internet connection. This module is used as a supplement in the course: Introduction to Engineering.

Introduction to Engineering (ENGR1100) is a 3-credit hour freshmen engineering course, in which students are introduced to the engineering process from problem formulation to the evolution of creative design. The objectives of this course are to excite students about engineering, cultivate problem-solving skills, encourage creativity, cultivate professionalism and emphasize the importance of communication skills. One approach to meeting these objectives is through the use of mechanical dissection/assembly activities.

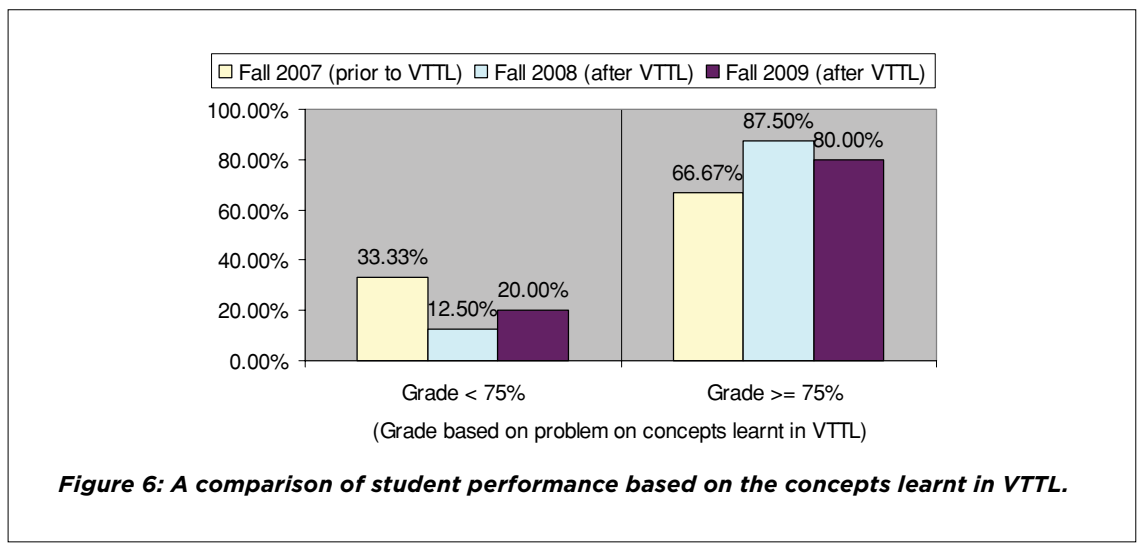

*Figure 6: A comparison of student performance based on the concepts learnt in VTTL.*

This mechanical dissection module in VIEW consists of a 3D CAD model of a mechanical power toothbrush modified using SolidWorks and imported into the virtual scene of the simulator using X3D. The models were originally created as part of the Cyber-Infrastructure-Based Engineering Repositories for Undergraduates (CIBER-U) project [28]. The graphical user interface of the simulator as shown in Figure 7 features HTML controls, a virtual 3D scene, and a control panel with a timer and scoring scheme. The scoring scheme and timer were used to emulate a gaming scenario as much evidence exists to support the effectiveness of digital game-based, interactive, student-learning environments [29]–[32]. The following URL: http://www.youtube.com/watch?v=pQWNjZqTnRE shows the simulator in use with the timer and scoring scheme. The module was implemented as team projects in ENGR1100 - the projects consisted of four phases, which were based on the engineering design process also introduced in this course [13].

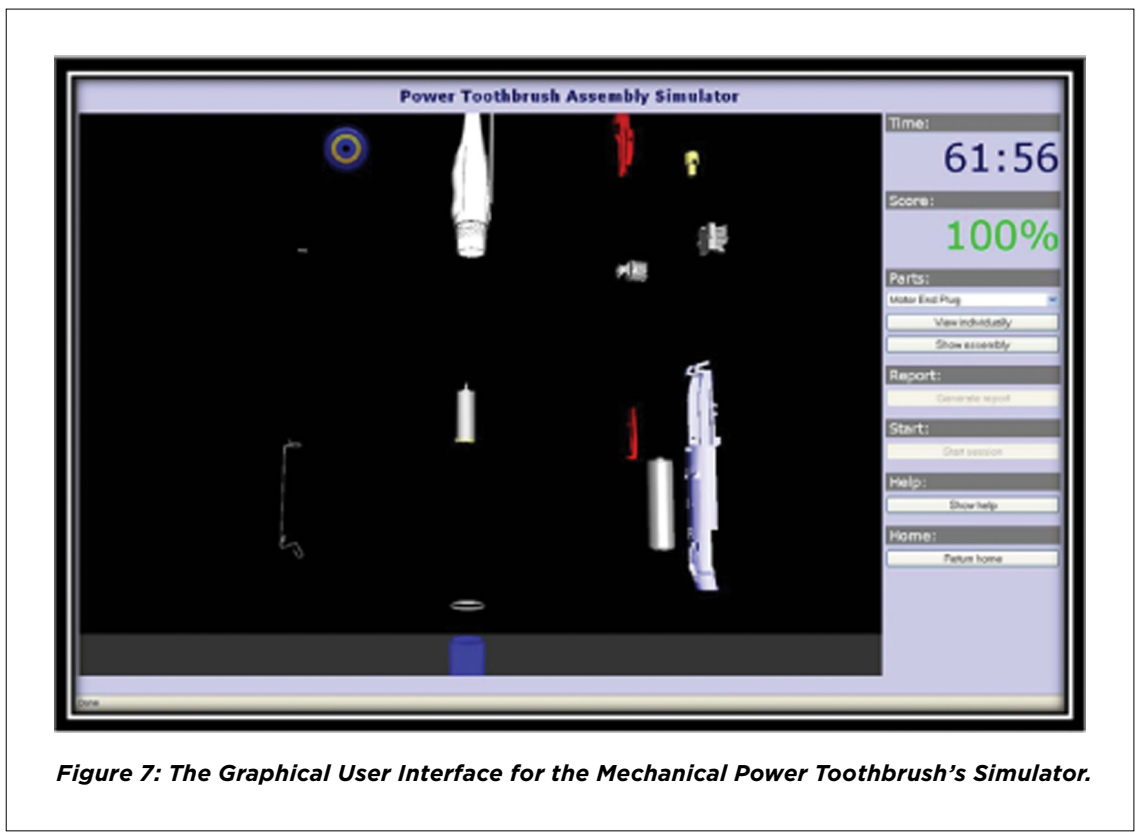

*Figure 7: The Graphical User Interface for the Mechanical Power Toothbrush's Simulator.*

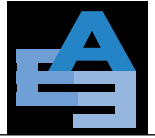

**Assessment**

Similar to the VTTL, the student perception survey results for the mechanical dissection module show that 97% or more of the students strongly agreed or agreed that the 3D models helped them to better visualize and understand the relationship between mechanical dissection and design. The following Figures 8–10 show a summary of the results from the surveys in Spring 2009, Fall 2009 and Spring 2010.

The bar graphs in Figures 8–10 correspond to the answers to the following questions:

**Question 1** (category Q 1): The use of interactive, 3D models in current engineering courses will help me to better understand the course material.

**Question 2** (category Q 2): I will gain more insight into the relationship between mechanical dissection and realize the design intent of each component by drawing, editing and grouping objects with an interactive 3D software program.

**Question 3** (category Q 3): The use of virtual engineering labs/modules will help to supplement the need for more 'hands on' projects and labs in the course.

**Question 4** (category Q 4): Introducing virtual engineering labs/modules will only make the curriculum more complicated.

**Question 5** (category Q 5): The 'computer game' scenario used in this module helps me to evaluate and learn about the various components used in the overall design of the product.

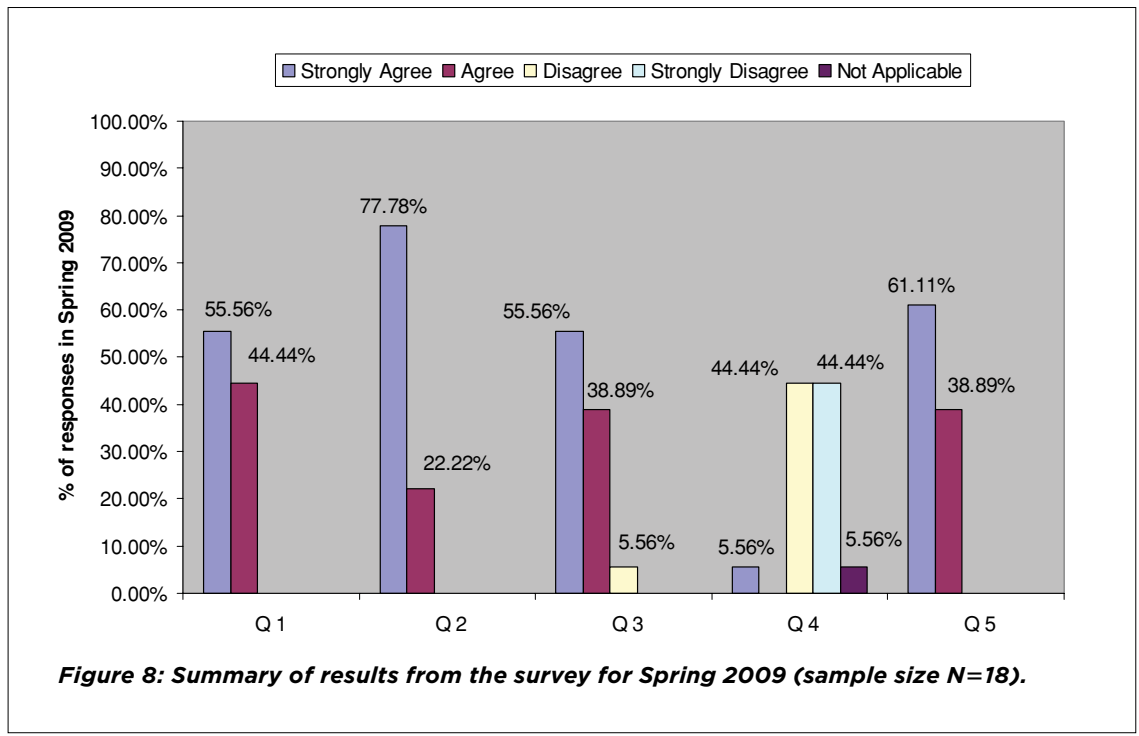

***Figure 8: Summary of results from the survey for Spring 2009 (sample size N=18).***

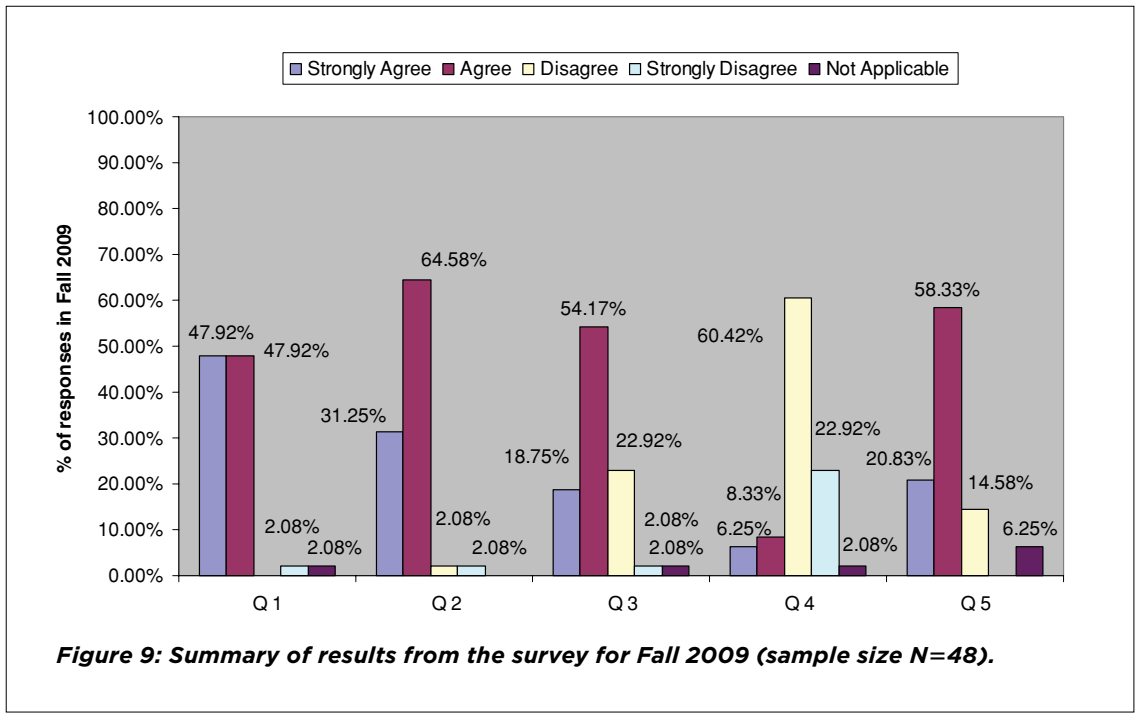

*Figure 9: Summary of results from the survey for Fall 2009 (sample size N=48).*

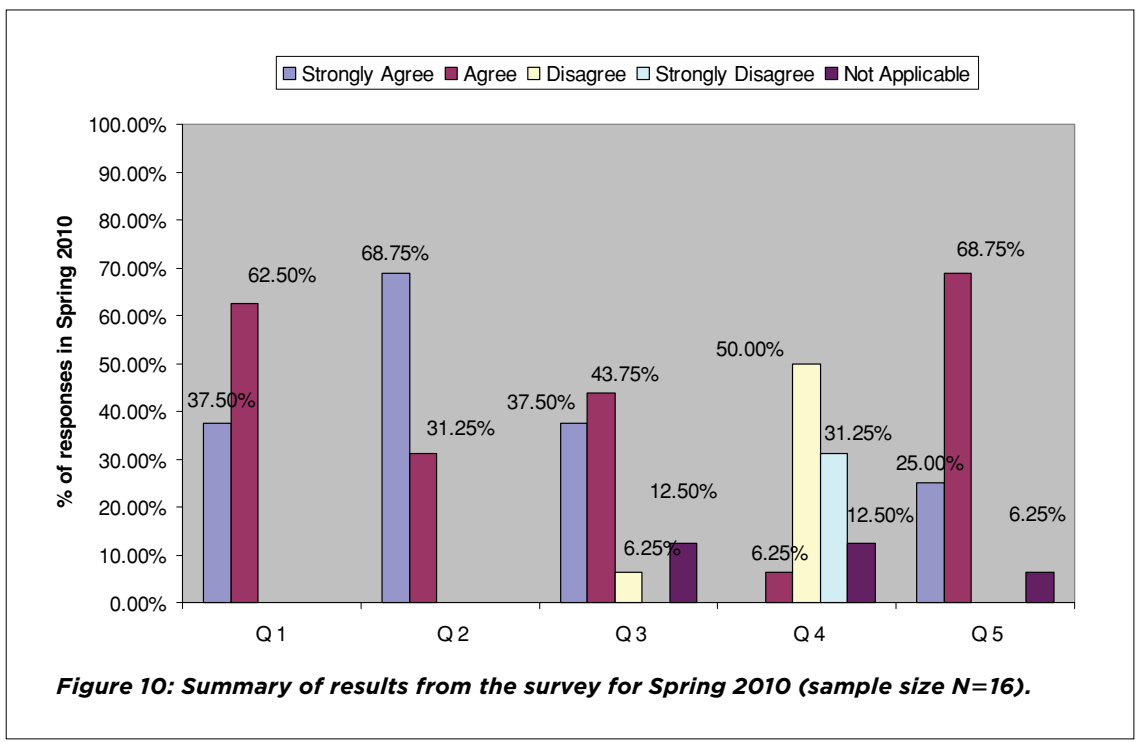

*Figure 10: Summary of results from the survey for Spring 2010 (sample size N=16).*

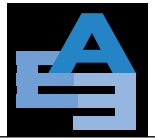

It is observed from the survey results shown in Figures 8–10 that 100% of the respondents in Spring 2009 and Spring 2010 and 96% of the respondents in Fall 2009 strongly agreed or agreed that the use of interactive 3D models in engineering courses would help them better understand course material. The 'computer game' scenario used in the simulator was also received well by majority of the students. Furthermore, the survey results show that the overall response was only 5.56% negative (1 out of 18 students; questions 3 and 4) in Spring 2009, 2.08%-14.58% negative (1–7 out of 48 students) in Fall 2009 with about 25% indicating that the use of virtual labs/modules will NOT help to supplement the need for more 'hands on' projects and labs in the course (question 3) and only 6.25% negative (1 out of 16 students, questions 3 and 4) in Spring 2010. This disparity in the overall responses between the spring and the fall semesters can be attributed to the general diversity in academic backgrounds (especially in this course), in the student body between the spring and fall semesters. The students enrolled in ENGR1100 in the spring semesters had an overall lower academic standing (based on GPA, SAT scores and the level of math/science completed) as compared with those in the fall semester. The disparity was amplified by the fact that 75% of the students enrolled in this course in Fall 2009 were GT Alliance students (dual enrolled at Georgia Institute of Technology and AASU), compared with only 5% of the same in Spring 2009 and Spring 2010. Further assessment in subsequent semesters would be useful to identify any relevant trends. In order to address this issue, the authors propose to develop other modules with relatively complex mechanisms and devices to challenge and engage students with different backgrounds. In addition, it is noted here that these virtual modules are designed to be used only as supplements in certain freshmen courses. Students are exposed to physical 'hands on' mechanical dissection activities later in the curriculum in courses such as Creative Decisions and Design (ENGR2110).

In addition to the surveys, the design schematics and descriptions submitted after every phase were evaluated to assess the effectiveness of the module on student learning. It was observed that the students developed a good understanding of the functions and roles of all the parts in the final product, and were extremely creative in their designs with an emphasis on crucial details important in such a product.

### PHYSICAL DISSECTION VS. VIRTUAL DISSECTION: A COMPARISON

There is still significant debate within the engineering education community pertaining to the effectiveness of virtual vs. physical dissection paradigms. In this section the authors address this issue and present the findings gained in a product dissection and assembly experiment carried out in both a physical and a virtual setting. In this context, the questions posed are:

1. Is it possible to determine the functionality and assembly sequence of the components of a mechanical device using only a virtual environment?
2. Is there a difference in terms of student learning and performance if this experiment is done in a traditional, physical setting?

One of the objectives of this research is to investigate whether Virtual Learning Environments (VLEs), such as VIEW, are suitable to provide students with a learning experience equivalent to that of a traditional in-class setting. In order to obtain data and investigate this problem, a test project was carried out at AASU in Spring 2010. As alluded to before (see section on the Mechanical Assembly of a Power Toothbrush), mechanical dissection and assembly experiments are an important component of many undergraduate engineering courses offered across the country. Due to a significant increase in distance learning programs as well as for a number of other reasons discussed earlier in this paper, more and more universities are undertaking efforts to allow students to conduct such experiments in virtual settings [33], [34]. In this realm, the intention of the authors is to compare the performance of students conducting such experiments in a physical hands-on laboratory with that of students who conducted the same experiment in a 3D Virtual Learning Environment: VIEW that was specifically developed for such purposes [13]. An overview of this experiment along with the insights gained is presented henceforth.

**Learning Objectives**

The student learning objectives for this exercise consisted of the following:

1. Determine the functionality of various components of a mechanical device (powered toothbrush) based on a virtual or physical examination of each component.
2. Reassemble the pre-dissected mechanical device using the proper sequence of the components.
3. Increase exposure to CAD modeling environment and applications of CAD models.

**Experimental Setup and Procedure**

The experiment was conducted in Spring 2010 using the same powered mechanical toothbrush discussed earlier (section on the Mechanical Assembly of a Power Toothbrush). It was completed by 23 students enrolled in the course ENGR1170 – Engineering Graphics who were assigned to two groups. Group A (11 students) conducted the experiment in a traditional, physical way. Group B (12 students) conducted the experiment in the VIEW Virtual Learning Environment. The students were selected to maintain a similar average GPA between the two groups; Group A students had a mean GPA of 3.18 (0.493 standard deviation), and Group B students had a mean GPA of 3.09 (0.744 standard deviation). The following steps of the experimental procedure were identical for both groups.

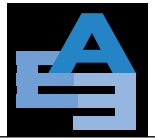

**Step 1**: Initially, the students in groups A and B were provided with the disassembled components of the toothbrush – either the physical parts or their 3D counterparts - and asked to describe the purpose and functionality of the individual parts. In order to measure their performance, a corresponding pre-assembly test was completed by the students within a given time frame of 10 minutes. A copy of the test can be obtained upon request from the authors.

**Step 2**: After the pre-assembly test, students were given 30 minutes to assemble the dissected components, ideally into a complete and fully functional device. The students in group A assembled the physical components of the toothbrush; the students in group B used the VIEW module to assemble the virtual components of the toothbrush.

**Step 3**: Next, all students were asked to complete the post-assembly test I which was identical to the pre-assembly test from step 1, however, this time with the experience of having assembled the individual components of the toothbrush into a complete device. The students were allowed to spend 15 minutes on this test. The purpose of this step was to test if their performance had improved based on their learning experience via the assembling process in step 2.

**Step 4**: Finally, the students were given a disassembled view (2D photographs in group A and 2D screen shots in group B) of the dissected components, and asked to show the correct order and position of assembly. This was referred to as the post-assembly test II, and students were given 10 minutes to complete this test. The purpose of this step was to test if the students were able to apply what they had learned from their hands-on physical or hands-on virtual assembly experiments, i.e., were they able to identify the correct assembly sequence based on the limited information provided and were there any differences in the performance of the virtual vs. physical groups?

**Assessment of Student Performance**

Both the results from the completed pre-assembly test (step 1) and post-assembly test I (step 3) were transcribed and the students' performance assessed based on a specific assessment rubric (shown in the appendix) that was developed for this purpose [35], [36]. Data were collected for both groups, i.e., for the physical experiment and the virtual counterpart.

The focus of the pre-assembly test and the post-assembly test I was on correct and accurate descriptions of the role/functionality of the various parts of the disassembled toothbrush. As for step 2, the time needed to assemble the toothbrush was measured. One point was also deducted for every mistake in the order or sequencing of the assembling process for both groups. It is noted here that the virtual module in VIEW (used by group B) was developed such that these points were deducted in real time, i.e., as soon as they occurred during the assembly process of step 2. However, it was not feasible to monitor the errors in real time during the physical assembly (group A). Thus, group A had points deducted based on an inspection of the assembled device at the end of step 2.

A qualitative comparison of the completed pre-assembly and post-assembly tests (see steps 1 and 3 above) showed that both groups' descriptions of the roles and functionalities of the dissected parts improved in terms of details and accuracy. This result was expected based on the assembling process in step 2, in which the students assembled the various components of the device either physically or virtually. This step provided both groups with greater insight as to the function and role of each component. Both groups were given 15 minutes for the post-tests instead of 10 minutes (as in the pre-test) to ensure greater detail in the post-test responses, which provided a more robust comparison between the post-test responses of the physical and virtual groups.

From a quantitative perspective, the authors compared assembly time and accuracy (step 2) and post-assembly test II (step 4) for both groups (see Figures 11–13 below). As shown in Figure 11, most of group A either improved or maintained their score on the post-assembly test II compared to the group's physical assembling process. Figure 12 shows there was a greater increase in the group B scores on the virtual post-assembly test II compared to the same groups' virtual assembling process. The average score for step 2 (physical or virtual assembling) for group A was 99% and 88.42% for group B. The higher score on the physical assembling exercise is a consequence of points deducted at the completion of step 2 compared to points deducted throughout the virtual assembling process as mentioned earlier. During the experiment it was observed that several group A students were making errors (without penalty) during the physical assembly exercise, but ultimately resolving these errors through trial and error and successfully completing the assembly. The average post-assembly test II scores were 99% and 98.67%, respectively for groups A and B. This result shows that both groups gleaned a similar level of understanding of how the components of the toothbrush fit together regardless of the physical or virtual environment used for steps 1–3.

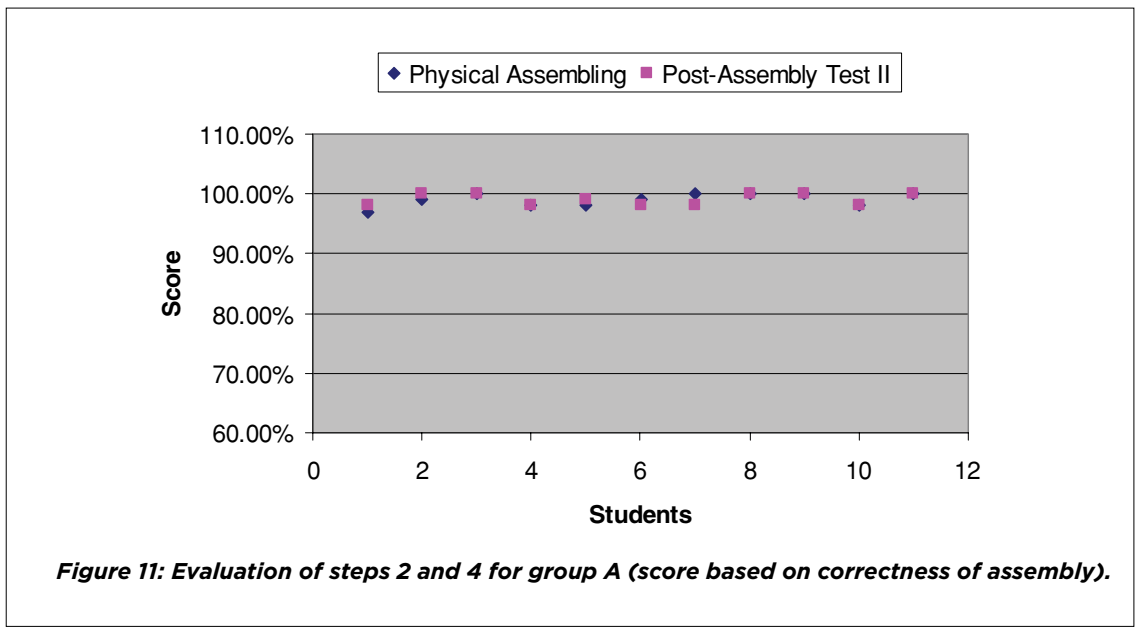

*Figure 11: Evaluation of steps 2 and 4 for group A (score based on correctness of assembly).*

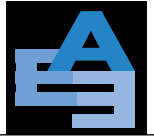

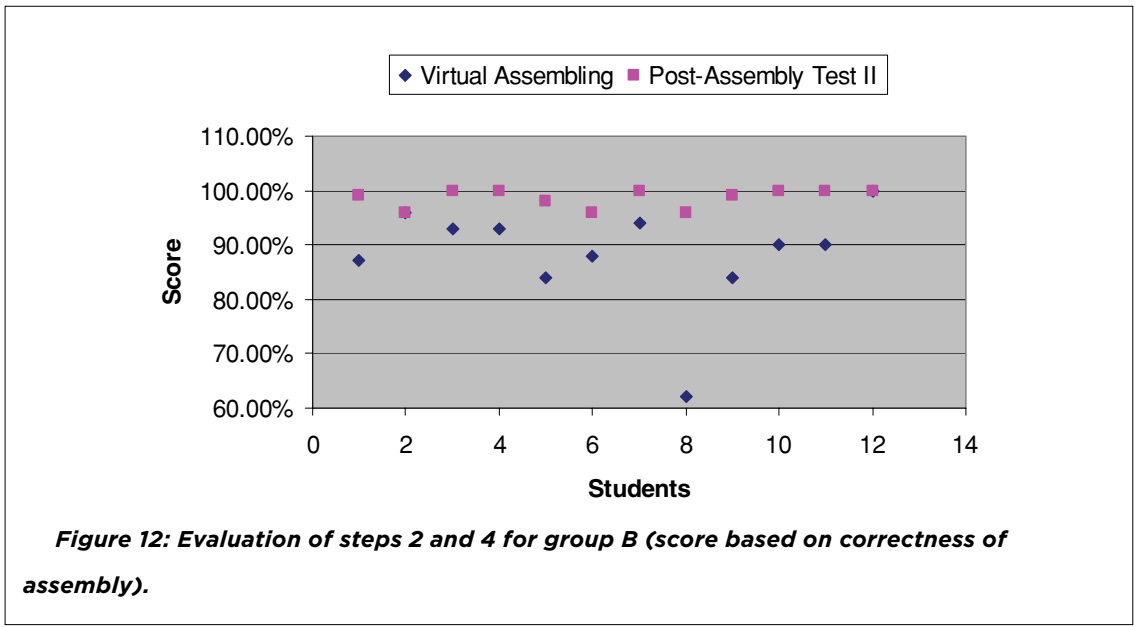

*Figure 12: Evaluation of steps 2 and 4 for group B (score based on correctness of assembly).*

Figure 13 shows a significant difference between the assembly times (step 2) for groups A and B. On average, the time needed to physically assemble the toothbrush was almost five times as long as the time needed to assemble the toothbrush in the virtual environment. This is due to the quasi-guided nature of the virtual environment. For example, in VIEW, the user is prevented (by the simulator) from assembling the wrong component in the assembly sequence in addition to a penalty for the mistake, where as, in the physical scenario, the user can continue to assemble the components and not realize an error in the sequence until later in time.

**Discussions on the Comparison**

As presented in the previous section, the responses and scores for the pre-assembly and post-assembly tests for both groups are identical with corresponding improvement in student performance from the pre-assembly to the post-assembly tests. Hence, it can be concluded that both groups fully met the desired learning objectives. This comparison of the performance of students conducting the dissection and assembly experiment in a traditional, physical way to that of students conducting the same experiment in a 3D Virtual Learning Environment reveals that: 1) It is possible to determine the functionality and assembly sequence of the components of a mechanical device using only a virtual environment, and 2) there is limited difference in student learning and performance for the same experiment in a traditional, physical setting.

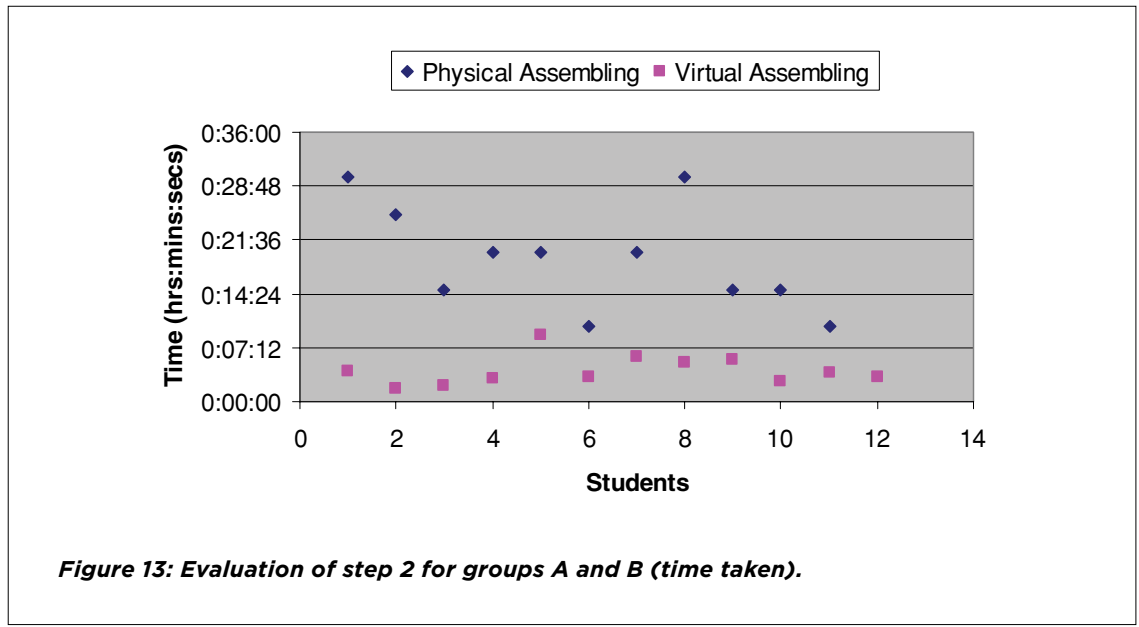

*Figure 13: Evaluation of step 2 for groups A and B (time taken).*

Overall, the findings agree with those of other researchers in the field. Recently, Hyder [37] com-pared the effectiveness of a number of different physical mechanical engineering undergraduate laboratory experiments with that of corresponding remotely controlled, web-based experiments and came to the conclusion that both can be considered equivalent if the web-based experiment is administered appropriately from a pedagogical point of view. Similar findings regarding the at-tainment of learning outcomes in virtual vs. physical learning environments are reported by Bright et al. [38] and Paretti et al. [39].

In addition to the above comparisons, it was also observed that in preparing the physical models for the assembling process, the toothbrushes had to be destructively disassembled e.g., the soldering for the battery connector had to be destructively removed, and it was not possible to fully disas-semble the brush subassembly on the toothbrush without completely destroying the components. This exemplifies how a virtual environment can be more flexible and beneficial than a correspond-ing physical dissection activity. This advantage is further bolstered with the consideration of the dissection of more complex devices and systems.

## CONCLUDING REMARKS AND FUTURE WORK

In this paper, the authors present a summary of the recent developments in project VIEW. The results of student surveys, comments and student performance show that VIEW has been beneficial to and well received by students. In addition, an experiment was performed to compare the learning experience and performance of students between physical and virtual dissection activities. The data from this study clearly suggests that for the given dissection and assembly experiment the learning experience and student performance in a physical vs. virtual learning environment can be considered equivalent. In order to further strengthen this conclusion, similar test projects will continue to be conducted for all modules available in VIEW.

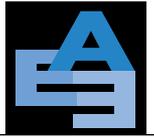

This work also lays the foundation for the development of additional virtual modules for these and other courses. For example, the development of the 3D CAD models for this project will also be presented as an application case study in the course Engineering Graphics (ENGR 1170), in which students learn SolidWorks, and a modified mechanical dissection module can be developed for the course Creative Decisions and Design (ENGR 2110). A new phase can also be developed for Engineering Graphics to help students visualize the spatial rotations of various geometrical solids. Other virtual laboratories used to evaluate material properties such as the flexural bending test, the Charpy impact testing, etc. can also be developed to supplement the Introduction to Engineering Materials (ENGR2000) and other courses. Based on assessment results and subsequent improvements to VIEW, it is proposed that these phases will be freely available to the general engineering education community.

## ACKNOWLEDGEMENTS

This work was supported by the AASU Teaching and Learning Grant 2007–2008 and the MATH + SCIENCE = SUCCESS AASU STEM Initiative Grants Program 2008–2009. The authors would also like to acknowledge support from Professor David Scott, Civil and Environmental Engineering, Georgia Institute of Technology, Savannah, GA for providing access to the equipment and required data. The work done by computer science students Ivan Sopin and Michael Brundage from the Network-Enabled Work Spaces (NEWS) research laboratory, AASU and engineering students Carlos Sanchez, Patrick Hager and Matthew Carroll is also appreciated by the authors.

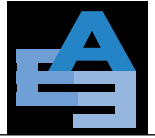

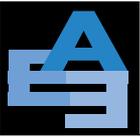

## AUTHORS

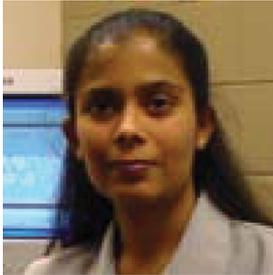

**Priya T. Goeser** is an Associate Professor of Engineering Studies at Armstrong Atlantic State University, Savannah GA. She received her Ph.D. in mechanical engineering from the University of Delaware and her B.Tech in mechanical engineering from the Indian Institute of Technology, Chennai, India. She also worked as a research associate at the Center for Composite Materials, University of Delaware prior to joining Armstrong Atlantic State University in 2003. Her current research interests are structural health monitoring, functionally graded materials, computational biomechanics and innovative teaching methods in engineering education.

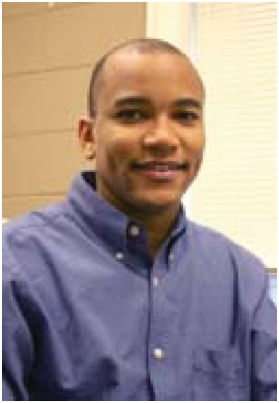

**Wayne M. Johnson** is an Associate Professor of Engineering Studies at Armstrong Atlantic State University in Savannah, GA. He received his Ph.D. in mechanical engineering from the Georgia Institute of Technology. His current research interests include mechatronics, vibrations and engineering education.

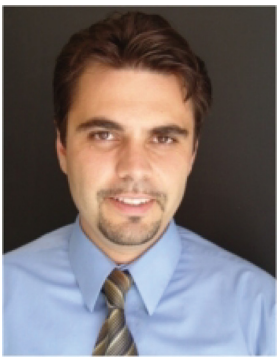

**Felix G. Hamza-Lup** is an Assistant Professor of Computer Science and Information Technology at the Armstrong Atlantic State University in Savannah, GA. He received a Ph. D. in Computer Science from University of Central Florida and a B.Sc. in Computer Science from Technical University of Cluj-Napoca, Romania. Felix's current research interests include human computer interaction, specifically multimodal interfaces with haptic capabilities, human perception, and distributed systems, specifically distributed simulation and web based e-learning systems.

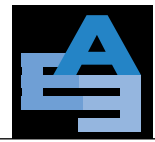

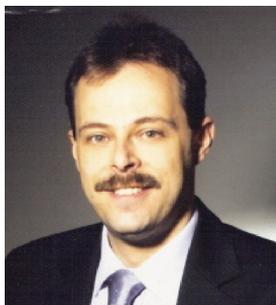

**Dirk Schaefer** is an Assistant Professor at the George W. Woodruff School of Mechanical Engineering at Georgia Institute of Technology. Prior to joining Georgia Tech, Dr. Schaefer was a Lecturer in the School of Engineering at Durham University, UK. During his time at Durham, he earned a Postgraduate Certificate in "Teaching and Learning in Higher Education" (PG-Cert). He joined Durham from a Senior Research Associate position at the University of Stuttgart, Germany, where he earned his Ph.D. in Computer Science. Dr. Schaefer has published more than 90 technical papers in journals, books and conference proceedings on Computer-Aided Engineering and Design as well as Engineering Education. Dr. Schaefer is a registered professional European Engineer (Eur Ing), a Chartered Engineer (CEng), a Chartered IT-Professional (CITP), a Fellow of the Higher Education Academy (FHEA) in the UK, and a registered International Engineering Educator (Ing-Paed IGIP).

**APPENDIX**

A sample of the rubric used for evaluation of the pre-assembly, assembling process and post-assembly tests is shown below.

| Date: | Assessment Rubric for Dissection / Assembly Experiment | | | | |
| --- | --- | --- | --- | --- | --- |
| Student: | | Group: | | Course: | |
| Activity/Assessment | poor | fair | average | very good | excellent |
| **Step 1 – Pre-assembly test: examination of dissected parts to understand their functionality and role as a part of the whole.** | *Little or no obvious understanding of role and functionality evident.* | *Obvious significant omissions in description of role, functionality AND/OR too much emphasis on one particular section.* | *Some omissions with regard to role, functionality or interaction with associated parts, or too much emphasis on one particular aspect.* | *Few omissions with regard to role, functionality with associated parts, or minor inaccuracies.* | *Role, functionality, and interaction with associated parts fully and accurately explained with details.* |
| **Step 2 – Assembling process: ability to correctly assemble given parts into a complete product** | *Not able to assemble parts in given time* | *Some parts correctly assembled, many errors in sequence.* | *Some/most parts correctly assembled; some understanding of interaction between parts* | *Most parts assembled in correct sequence, very good understanding of interaction between parts* | *All parts assembled in correct sequence without any errors* |
| **Step 3 – Post-assembly test I: examination of dissected parts to understand their functionality and role as a part of the whole, after assembly experience.** | *Little or no obvious understanding of role and functionality evident.* | *Obvious significant omissions in description of role, functionality AND/OR too much emphasis on one particular section.* | *Some omissions with regard to role, functionality or interaction with associated parts, or too much emphasis on one particular aspect.* | *Few omissions with regard to role, functionality or interaction with associated parts, or minor inaccuracies.* | *Role, functionality, and interaction with associated parts fully and accurately explained with details.* |
| **Step 4: Post-assembly test II: Ability to determine correct assembly sequence based on 2D information.** | *Not able to determine valid assembly sequence* | *Some parts connected in correct sequence* | *Most parts connected in correct sequence* | *Assembly sequence correctly determined with some errors.* | *Assembly sequence correctly determined without any errors.* |